\documentclass[a4paper,11pt]{article}
\usepackage{xcolor}
\pdfoutput=1 

\usepackage{jcappub} 

\usepackage{hyperref}
\usepackage{comment}

\usepackage[T1]{fontenc} 

\title{\boldmath Looking for Black Hole Morsels in Astrophysical Mergers via Hawking Radiation}

\author[a,b,1]{Giacomo Cacciapaglia,\note{Previously at Universite Claude Bernard Lyon 1, CNRS/IN2P3, IP2I UMR 5822, 4 rue Enrico Fermi, F-69100 Villeurbanne, France.}}
\author[b,c,d]{Stefan Hohenegger,}
\author[b,d,e,f]{Francesco Sannino}

\affiliation[a]{Laboratoire de Physique Th\'eorique et Hautes \'Energies (LPTHE), UMR 7589,
Sorbonne Universit\'e \& CNRS, 4 place Jussieu, 75252 Paris Cedex 05, France}
\affiliation[b]{Quantum Theory Center ($\hslash$QTC) at IMADA \& D-IAS, Southern Denmark Univ., Campusvej 55, 5230 Odense M, Denmark}
\affiliation[c]{Universite Claude Bernard Lyon 1, CNRS/IN2P3, IP2I UMR 5822, 4 rue Enrico Fermi, F-69100 Villeurbanne, France}
\affiliation[d]{Dept. of Physics E. Pancini, Universit\`a di Napoli Federico II, via Cintia, 80126 Napoli, Italy}
\affiliation[e]{INFN sezione di Napoli, via Cintia, 80126 Napoli, Italy}
\affiliation[f]{Scuola Superiore Meridionale, Largo S. Marcellino, 10, 80138 Napoli, Italy}

\emailAdd{cacciapa@lpthe.jussieu.fr}
\emailAdd{s.hohenegger@ipnl.in2p3.fr}
\emailAdd{sannino@qtc.sdu.dk}

\abstract{Gravitational wave observation has provided numerous insights into the merger of astrophysical black holes. In contrast to other violent events (\emph{e.g.} supernovae), they are, however, not expected to lead to significant emissions of photons and neutrinos. In this paper we discuss a scenario that would lead to characteristic observable gamma ray bursts, which would provide numerous hints to physics beyond General Relativity. Starting from the hypothesis that micro-black holes (called \emph{morsels}) are formed during the merger process, we show that it is possible to observe their Hawking radiation, which takes the form of gamma ray bursts of a uniquely characteristic form: with energies in the TeV range, their temporal structure is unlike that stemming from any other astrophysical event. Notably, the time delay from the gravitational wave event is correlated to the mass distribution of the morsels. The integrated mass of the morsels, allowed by the unaccounted merger mass, leads to a Hawking radiation in photons that is above the sensitivity of atmospheric Cherenkov telescopes such as HESS, LHAASO and HAWC, and gamma ray space telescopes, such as Fermi-LAT. This renders the hypothesis of morsel creation experimentally testable, and we provide the first concrete bounds on the total mass of morsels formed in specific events.}

\begin{document}
\maketitle
\flushbottom

\section{Introduction}

Black Holes (BHs) are extreme objects predicted by Einstein's theory of General Relativity (GR), whose existence has been demonstrated experimentally in numerous different ways: due to their nature, most detections are indirect, \emph{i.e.} BHs reveal themselves through their gravitational attraction on other (directly observable) objects. In this way, it is by now well established that super-massive Black Holes (SMBHs) of $10^5-10^9$ solar masses are present in the centre of most galaxies \cite{Kormendy:1995er,Richstone:1998ky}, where they significantly affect the rotation speed of nearby stars \cite{Ghez:1998ph,Schodel:2002py}. 
Recent advances in large-scale astrophysical experiments, however, have provided new and more direct ways of detecting and measuring BHs. For example, SMBHs can form active galactic nuclei and a first direct image of such a space-time region has been taken in \cite{EventHorizonTelescope:2019dse}. Furthermore, stellar mass BHs are produced at the end of heavy star lifetimes, and their mergers are at the origin of the gravitational wave signals observed at LIGO/VIRGO interferometers \cite{LIGOScientific:2016aoc}. 

Further, even more direct ways of directly observing BHs have been theorised within frameworks going beyond GR: concretely, by incorporating quantum effects, a distinctive signal from BHs may be due to the emission of particles via the evaporation mechanism introduced by S.~Hawking  \cite{Hawking:1974rv}:  based on semi-classical computations, it was argued in this seminal work that BHs have thermodynamic properties and behave like (perfect) black bodies by emitting thermal radiation. A direct observation of this \emph{Hawking radiation}\footnote{Hawking radiation has been experimentally observed in analogue BHs in table-top experiments, see for instance \cite{Steinhauer2014,Modugno2021,Kolobov2021}.} would give us invaluable insights not only into properties of BHs but also into possible theories of quantum gravity beyond GR. However, it is extremely challenging due to a number of different aspects: 
\begin{enumerate}
\item[\emph{(i)}] The effective temperature is inversely proportional to the BH mass. Hence  astrophysical BHs heavier than a few solar masses cannot emit Hawking radiation, because they are colder than the cosmic microwave background. Only the radiation stemming from BHs lighter than half the moon mass, $\sim 3\times 10^{22}$~kg, can be hoped to be observable.
\item[\emph{(ii)}] Due to its nature as a black body radiation, it is not directional. Therefore, the flux of Hawking radiation from a (single) distant BH arriving on earth is heavily diluted by the distance squared.
\item[\emph{(iii)}] Sufficiently small BHs emit radiation until they completely evaporate: this evaporation process accelerates as the mass diminishes, leading to signals of a burst-like characteristic (with high intensity and high radiation energies) towards the end of a BH lifetime.
\end{enumerate}
These properties suggest that experimentally detectable signals would require very peculiar configurations of BHs, namely \emph{small} BHs either in close proximity to earth or of sufficient multiplicity to generate a detectable flux. Small BHs weighing like  asteroids may have been produced in the early Universe \cite{Hawking1971PBH} within high density regions. However, these \emph{primordial BH}s would also have already evaporated during these early cosmological stages and the corresponding Hawking radiation is constrained by measuring the diffuse gamma ray background \cite{Kimura:2016xmc}.~\footnote{Assuming that primordial BHs are long-lived by some mechanisms, they have been discussed as potential candidates for dark matter \cite{Chapline:1975ojl,Meszaros:1975ef,Frampton:2010sw,Villanueva-Domingo:2021spv}.} 

While stellar-sized BHs are natural end points of the life-cycle of heavy stars, and even larger ones can be created through consecutive mergers or accretion by matter, possible creation mechanisms for smaller BHs, other than primordial ones, are far more speculative and typically invoke physics beyond GR or beyond the Standard Model of particle physics (see \emph{e.g.} \cite{Chitishvili:2021jrx}). In such theories, sufficiently energetic (and potentially catastrophic) astrophysical events (such as the merger of two astrophysical BHs) could produce, as a byproduct, a large number of small BHs. These BH debris, if sufficiently light, evaporate by emitting Hawking radiation and provide an electromagnetic and neutrino signal counterpart to the astrophysical event (such has gravitational waves from BH mergers). The observation of such an event would give us invaluable insights into quantum properties of gravity beyond GR, not only through the measurement of the Hawking radiation, but also by giving hints towards potential creation mechanisms of these small BHs themselves. Therefore, given the obvious inherent theoretical and experiment interest in such a signal, in this paper we shall ask the question: Can such an event be experimentally detected and  distinguished from other astrophysical signals? 

Concretely, we start from the \emph{hypothesis} that a significant number of small BHs, can be created during a catastrophic event such as the merger of two astrophysical BHs.\footnote{We shall provide additional motivation for this hypothesis in Appendix~\ref{App:Motivation}, by reviewing fragmentation processes in theories beyond GR. We also provide a very simple back-of-the-envelope estimation for rates and sizes of created BHs by invoking the central dogma~\cite{Strominger:1996sh}.} In order to distinguish these from the `large' BH created in the merger (and also primordial BHs), we shall call these \emph{black morsels}. We thus explore the observational consequences of the production of such morsels during BH mergers: while mergers can be detected via gravitational waves, they are generally not expected to be accompanied by gamma ray bursts (GRB) or high-energy neutrino emission. An electromagnetic counterpart, in fact, is contingent on the presence of pre-existing material from their progenitor star \cite{Loeb_2016,Woosley_2016}, from supernova remnants \cite{Perna:2016jqh,Kimura:2016xmc,Murase_2016} or from active galactic nuclei \cite{Bartos_2017}. However, as we shall show in this paper, the creation of morsels (and their subsequent evaporation due to Hawking radiation) would lead to a signal of high-energetic photons and neutrinos that carries two marked features:
\begin{itemize}
\item Since the evaporation time of morsels depends on their mass and high energy particles are only emitted towards the end, the onset of the visible gamma ray signal could be significantly delayed with respect to the arrival of the gravitational waves. This delay is correlated to the mass distribution of the morsels. Additional effects may stem from the gravitational field of the merged BH, depending on  where the morsels were created.
\item The photon and neutrino energies stemming from  morsels typically exceed the TeV scale, unlike astrophysical sources. Moreover, since the wave-length of the Hawking radiation becomes smaller as the morsels approach the point of evaporation, the  signal is expected to rise in energy and intensity, before disappearing (when all morsels have evaporated).
\end{itemize}
We thus argue that the Hawking radiation stemming from these BH morsels gives rise to GRBs possessing a distinctive fingerprint and with energy flux observable at current gamma ray telescopes: this therefore allows for an experimental verification of our hypothesis (\emph{i.e.} that morsels are produced in BH mergers) at current atmospheric Cherenkov telescopes, such as HESS, LHAASO and HAWC, which are capable of measuring TeV GRBs, and gamma ray space telescopes, such as Fermi-LAT, in the multi-GeV energy range. The uncertainties from gravitational wave detection of BH mergers still allow the emission of a substantial mass (of the order of solar masses) in black morsel debris. Assuming a simple mass distribution of the morsels (motivated by the central dogma~\cite{Strominger:1996sh}), we provide a concrete form of the photon signal that would be experimentally detected on earth. By comparing with actual experimental data, we provide the first upper limits on the total mass emitted in the form of morsels during BH mergers.

We remark that the possibility of detecting electromagnetic radiation from evaporating black holes has been previously suggested in \cite{Chitishvili:2021jrx} (see also \cite{Cai:2021zxo} for earlier work). However, the authors consider a specific production mechanism (following a merger), which leads to  short-lived micro-BHs and, thus, to the rapid emission of high energy photons and neutrinos (typically below one second), which would be observed on earth during the arrival of the GW signal of the merger or shortly thereafter. In the current work we consider larger morsel masses, which generate slower Hawking radiation emission and signals observable over longer time-scales. This leads to a \emph{detectable} signal over much longer observation times, which we explicitly calculate and characterize.

While the idea of BH fragmentation during mergers has not, to our knowledge, been directly addressed in the literature, there exists extensive work on post-merger phenomena leading to multimessenger signals. For example, mergers of compact objects could lead to gravitational wave “kicks”~\cite{Campanelli:2007cga, Lousto:2012su}, accretion disk disruption~\cite{Rosswog:2005su, Metzger:2011bv}, and transient electromagnetic phenomena~\cite{LIGOScientific:2017ync}. Our scenario differs fundamentally in that it hypothesizes the formation of genuinely distinct compact objects bound by a horizon. We now make this novelty explicit and provide more details on the possible morsel formation in Appendix~\ref{App:Motivation}.

\section{Observable signatures}

Since we consider the creation of BH morsels as a (testable) hypothesis, we assume shall assume a distribution of morsel masses that escapes the merger (see also Appendix~\ref{Sect:Estimation} for a back-of-the-envelop estimate based on the central dogma). In other words, when the two BHs merge into a single one, regions of space-time characterised by a strong non-linear gravitational field lead to the formation of  BH morsels outside the final horizon. The physics of the Hawking radiation of the BH morsels is encoded in their mass distribution function 
\begin{equation}
  \rho(M_{\rm Bm}) \equiv \frac{d n_{\rm Bm}}{d M_{\rm Bm}}\,.
\end{equation}
Here, $\rho_{\rm Bm} \, d{M_{\rm Bm}}$ gives the number of BH morsels in the infinitesimal mass range between $M_{\rm Bm}$ and $M_{\rm Bm}+dM_{\rm Bm}$. For now, we neglect the effect of the gravitational field of the merged BH, and consider the morsels to be in a spherically-symmetric shell configuration. Each morsel emits particles with rates that depend on its mass. The emission rates per each particle species $p$ follow the master equation \cite{Hawking:1975vcx}
\begin{equation}
    \frac{d^2\, N_p}{dt \; dE_p} = \frac{1}{h} \frac{\Gamma_p (E_p, M_{\rm Bm})}{\exp \frac{E_p}{k_B T_{\rm Bm}} \pm 1}\,,
\end{equation}
where the denominator is the usual Boltzmann statistic factor for BH morsel temperature $T_\text{Bm}$ (with $\pm 1$ for fermions/bosons, respectively) and the numerator contains the specific grey body factors. Hence, the emission rate for the distribution of BH morsels can be computed as
\begin{equation}
    J_p =  \frac{1}{2} \int_{M_{\rm min}}^{M_{\rm max}} dM_{\rm Bm}\; \frac{dn_{\rm Bm}}{dM_{\rm Bm}}\; \frac{d^2\, N_p}{dt \; dE_p} (\mu_t (M_\text{Bm}, t))\,,
\end{equation}
where $\mu_t$ is the mass of the morsels at time $t$ as a function of the initial mass $M_\text{Bm}$. The geometric factor of $1/2$ counts the particles emitted away from the merged BH.
The above formula yields the differential primary flux for each particle species $p$. We computed the BH evolution and $J_p$ numerically by use of the open-source public code BlackHawk \cite{Arbey:2019mbc,Arbey:2021mbl}, where radiative emission, decays and hadronisation of the primary particles is taken into account by using Pythia \cite{Sjostrand:2014zea}.  
The hadronisation tables, therefore, allow us to reliably generate secondary radiation for primary particle energies between $5$~GeV up to a few TeV.
For the numerical results in this letter, we use as template a population of BH morsels with equal masses and non rotating, as angular momentum dissipates faster than the mass of the evaporating BH \cite{Page:1976ki}.

\subsection{Stellar mass mergers}

For LIGO/VIRGO/KAGRA mergers \cite{LIGOScientific:2018mvr,LIGOScientific:2020ibl,LIGOScientific:2021usb,KAGRA:2021vkt}, the BH masses are of stellar order, ranging from a few to several tens of solar masses, with associated distance span between 240 Mpc to 3 Gpc. The masses of the initial and final BHs can be indirectly measured from the gravitational wave spectrum, with errors of a few solar masses, depending on the event. For instance, for GW170814 \cite{LIGOScientific:2017ycc}, which is the first BH merger observed by all three detectors of LIGO and VIRGO, the initial BH masses have been determined to be $30.5^{+5.7}_{-3.0}$ and $25.3^{+2.8}_{-4.2}$, in units of the solar mass $M_\odot$. The final BH mass is $53.2^{+3.2}_{-2.7}\; M_\odot$, while the total energy emitted in gravitational waves amounts to $2.7^{+0.4}_{-0.3}\; M_\odot$. Taking into account the errors in the parameters, the energy budget of the merger still allows for a total energy of a few solar masses to be emitted in BH morsels. Conservatively, we will take one solar mass as a reference value for the maximal total mass in morsels.

\begin{figure}[tb!]
\centering
\includegraphics[width=7.5cm]{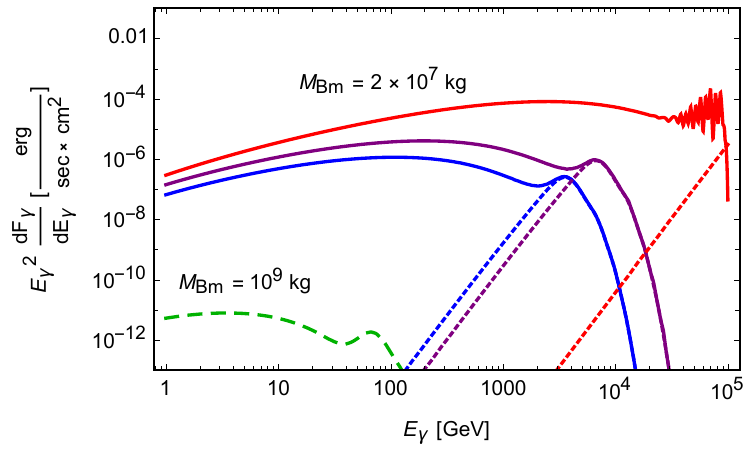} \hspace{0.3cm}\includegraphics[width=7.1cm]{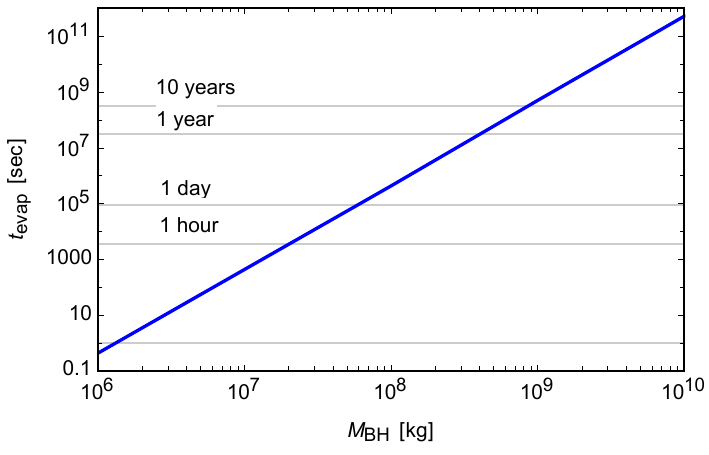} 
\caption{\label{fig:results} Left: Differential flux $E_\gamma^2 dF_\gamma/dE_\gamma$ emitted by a distribution of BH morsels of same mass at a distance of $300$ Mpc. The solid lines correspond to $M_{\text{Bm}}=2 \times 10^7$~kg at different times: until 500 sec (blue), at 3000 sec (purple) and at 3400 sec (red). The latter is close to the evaporation time. The dotted lines show the primary spectrum. For comparison, in green we show the spectrum for a mass of $10^9$ kg, for which the emission remains constant over a long time period. The fluxes are normalised to a total morsel mass of 1 $M_\odot$. Right: Evaporation time as a function of the BH mass, computed via BlackHawk.}
\end{figure}

Via evaporation, the BH morsels emit all kinds of SM particles, however only the neutral stable ones, photons and neutrinos, reach detectors on Earth without being deflected by galactic magnetic fields. To compute template photon and neutrino fluxes, we consider as benchmark a BH merger at a distance $D=300$~Mpc from Earth, corresponding to the range of the closest detected signals. The photon flux on Earth can be computed as
\begin{equation}
    F_\gamma = \frac{1}{4 \pi D^2} \int dE_\gamma\ J_\gamma\,.
\end{equation}
In the left panel of Fig.~\ref{fig:results} we show the differential fluxes, $E_\gamma^2 dF_\gamma/dE_\gamma$, for a distribution of BH morsels of same mass, normalised to a total of one $M_\odot$. The solid lines provide templates for morsel masses of $2 \times 10^7$~kg, where the colours indicate different times from the production. As the evaporation lifetime of such mass is about 1 hour, the emission remains constant for timescales of about $500$~sec, emitting the spectrum illustrated by the blue line. Towards the final evaporation time, both the emissivity and the particle energies increase, approaching the  `explosion' at the end of the BH lifetime \cite{Hawking:1974rv}. For the red curve, close to the evaporation time, the primary energies exceed $100$~TeV, reaching the current limit of the BlackHawk code with Pythia secondary modelling. In fact, the wiggles at high energies correspond to numerical instabilities in the code. We also included the effect of the optical transparency of the intergalactic medium, which cuts off the propagation of photons with energies exceeding $100$~TeV \cite{Cirelli:2010xx}, manifesting as a sudden drop in the red curve. For reference, the dotted lines show the primary emission of photons. 
For different morsel masses, the evaporation time will change drastically as it is proportional to $M_\text{Bm}^{3}$. Lighter BHs will, therefore, generate an energetic GRB in a shorter time. Instead, heavier morsels provide a steady signal at the beginning, characterised by reduced particle energies, proportionally to the Hawking temperature $T_\text{Bm} \propto M_\text{Bm}^{-1}$, while the overall luminosity is reduced. As a reference, we show in dashed green the early curve for a mass of $10^9$~kg, corresponding to an evaporation lifetime of about 16 years. Nevertheless, all initial BH morsel masses will generate an energetic GRB with similar characteristics with a time delay from the merger of the order of their evaporation time. In the right panel of Fig.~\ref{fig:results} we show the evaporation lifetime as a function of the BH mass, as computed by BlackHawk. Similar fluxes can be obtained for neutrinos.

\vspace{0.3cm}

\begin{figure}[tb!]
\centering
\includegraphics[width=8cm]{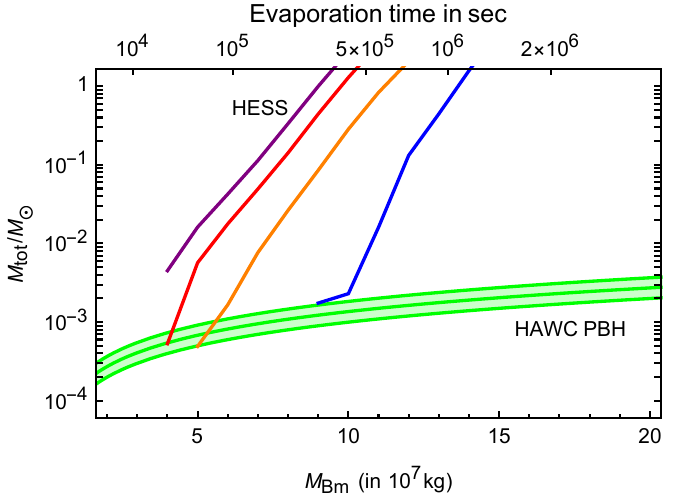} 
\caption{\label{fig:HESS} Estimated upper limit on the total mass $M_\text{tot}$ emitted in BH morsels as a function of the morsel mass from the HESS observation of four LIGO/VIRGO events \cite{HESS:2021gbx}. The curves are extracted from GW170814 (blue), GW190512$\_$180714 (red), GW190728$\_$064510 (orange) and GW200224 (purple). The upper labels indicate the evaporation time as a function of the BH morsel masses. The green band indicates a reinterpretation of the HAWC bound on primordial BHs \cite{HAWC:2019wla}.}
\end{figure}

Multimessenger astrophysics associated with gravitational wave events is part of the LIGO/ VIRGO/KAGRA programme, and electromagnetic counterparts are mainly being monitored by Fermi-GBM and Swift-BAT experiments within 30 seconds from all event alerts \cite{FermiGamma-RayBurstMonitorTeam:2023mtr}. Motivated by neutron star mergers, the coverage is between a few keV and MeV. As shown in Fig.~\ref{fig:results}, we expect very small signals from the BH morsels within this range, well below the current experimental sensitivity of  $10^{-7}$~erg sec$^{-1}$ cm$^{-2}$ for the flux on Earth. In 2016, however, a short GRB was associated to a BH merger event \cite{Perna:2016jqh,Connaughton:2016umz,Greiner:2016dsk}, prompting interest in exploring electromagnetic counterparts of BH mergers \cite{Perna:2019pzr}. The expected signals are mainly due to pre-existing material surrounding the BH binary system, hence triggering searches in the optical \cite{DES:2018uhh,Lundquist_2019,Turpin:2019bjj,DLT40:2019iur,Grado:2020zjc,Kim:2021hhl} or keV/MeV \cite{Ferrigno:2020lli,Klingler:2019fbl,Klingler:2020osq} ranges. The HESS telescope has followed four BH mergers from the O2 and O3 runs of LIGO/VIRGO \cite{HESS:2021gbx}, covering photon energies between 1 and 10 TeV. However, the observation could only focus on the region of the sky at a time $10^4 \div 10^5$ sec after the BH merger occurred. As no significant point-like source was found in the merger region of the sky, the energy flux from the BH mergers could be constrained to be below $10^{-12}~\text{erg}\ \text{sec}^{-1}\ \text{cm}^{-2}$. For morsel masses corresponding to evaporation times within or after the observation window, the HESS data can be used to extract a bound on the total mass that can be emitted as a function of the morsel masses. For each event, we compare the limit on the luminosity to the maximal luminosity of the morsel Hawking radiation within the respective observation window, computed by integrating the source emission between $1$ and $10$~TeV. The results are shown in Fig.~\ref{fig:HESS}, where the strongest bounds come from the closest events, GW170814 \cite{LIGOScientific:2017ycc} and GW190728$\_$064510. The BH morsel mass ranges stem from the different observation times, where masses smaller than the start of the curves are unconstrained as total evaporation occurs before the HESS observation window. For morsel masses above $15 \times 10^7$~kg, the energies of the emitted particles fall below the sensitivity range of HESS, and the constraint quickly worsens. This should be taken as a simple estimate, as the limits in \cite{HESS:2021gbx} were obtained for a $E^{-2}$ power-law spectrum of the photons, which does not exactly match the spectra in Fig.~\ref{fig:results}. 

As visible signals from BH evaporation always entail photons above the TeV energy, this signal offers a golden opportunity for high energy atmospheric Cherenkov telescopes like HESS, HAWC \cite{HAWC:2020hrt} and LHAASO \cite{LHAASO:2023kyg,LHAASO:2023rpg,LHAASO:2023lkv}. The latter recently published a catalogue of multi-TeV sources, reporting the observation of differential fluxes within the range $10^{-13} \div 10^{-11} ~\text{erg}\ \text{sec}^{-1}\ \text{cm}^{-2}$~\cite{LHAASO:2023rpg}. Note that, if the BH morsels have masses above $10^9$~kg and evaporation times above several years, BH mergers predating the first LIGO/VIRGO detection would leave behind a relic BH morsel shell and produce a TeV scale GRB uncorrelated to merger catalogues, similar to the signal studied in \cite{Cai:2021zxo}. Henceforth, a search for BH morsels could follow a similar strategy to searches for local densities of primordial BHs (PBHs). Currently, the strongest bound comes from HAWC \cite{HAWC:2019wla,HESS:2023zzd}. For morsels of same mass, the signal from a morsel distribution at a distance $D$ is equivalent to a single PBH at a distance $D_{\rm PBH} = D/\sqrt{n_{\rm Bm}}$. Hence, under the simplifying assumption that all BH mergers eject the same amount of morsels of the same mass, one can estimate the limit on the BH morsel fraction by comparing the rate of BH mergers measured by LIGO/VIRGO \cite{LIGOScientific:2020kqk}, $\rho_{\rm LV} = 24^{+14}_{-9}\;\mbox{Gpc}^{-3}\ \mbox{yr}^{-1}$, to the rescaled limit on PBH densities from HAWC, $\rho_{\rm PBH} = 3400\; \mbox{pc}^{-3}\ \mbox{yr}^{-1}$. The estimated lower limit on the total morsel mass is shown in Fig.~\ref{fig:HESS} in green, where the band comes from the uncertainty in the BH merger rate. A dedicated search, taking into account the morsel mass distributions for different merger events, can provide a more reliable limit.

Neutrino observatories like ANTARES and IceCube have also monitored BH mergers in search for neutrino counterparts \cite{ANTARES:2018bmu}. Both ANTARES \cite{ANTARES:2020kxp,ANTARES:2022gyy} and IceCube \cite{IceCube:2021ddq} search for neutrino events within a window of $\pm 500$~sec around the BH merger, finding no excess over the expected backgrounds. However, the limits in terms of fluxes are orders of magnitude above the ones for photons, requiring typical source luminosity of $10^{51}$~erg sec$^{-1}$. As the BH morsels predict similar fluxes in photons and neutrinos, searches for the latter remain not competitive.

\subsection{Uncertainties and Sensitivity Analysis}
The predictions presented in the previous section rely on numerical computations from BlackHawk and on assumptions regarding the distribution, number, and initial conditions of the BH morsels. Here, we recap the uncertainties stemming from:
\begin{itemize}
    \item \textbf{BlackHawk Limitations:} We note that the BlackHawk software used for our calculations relies on semiclassical assumptions and a fixed (non-backreacting) background. Hence, it does not capture possible quantum gravity or dynamical spacetime effects~\cite{Arbey:2019mbc,Arbey:2021mbl}. Another source of uncertainty stems from the hadronisation model, which typically affects the results at the $20$~\% level.
    \item \textbf{Morsel Mass Distribution:} We considered a single mass distribution, with typical morsel masses ranging from $10^7$ to $10^9$ kg in order to have an observable time-delay from the merger. Changing the  initial mass distribution would alter the predicted flux normalization and duration.  
    \item \textbf{Spin Effects:} While we assume non-rotating morsels, including initial spin decreases the Hawking temperature, shifts the spectrum to lower frequencies, and introduces anisotropic emission.  Furthermore, the evaporation is slowed down, however the spin dissipation is faster \cite{Page:1976ki}. Hence, including spin effects should lead to minor modifications of our results.
    \item \textbf{Merger Rate Uncertainties:} Published LIGO/VIRGO rates vary by an order of magnitude \cite{LIGOScientific:2020kqk}, directly affecting the expected detection rate for non-coincidence searches. This uncertainty has been included in the green curve in Fig.~\ref{fig:HESS}.
\end{itemize}
As a first estimate, we indicate the impact of parameter variation in Table~\ref{tab:uncertainty}. Future work should include full Monte Carlo parameter scans and systematic error propagation.

\begin{table}[h]
\centering
\begin{tabular}{|l|c|c|}
\hline
\textbf{Parameter} & \textbf{Variation} & \textbf{Flux Impact (approx.)} \\
\hline
Numerical (BlackHawk)       & hadronisation & $\sim 20$~\% \\
Morsel mass distr.       & $10^7$--$10^9$\,kg & Duration: min--yr; Flux amplitude: $\sim$10$\times$ \\
Initial spin       & $a_*=0$--$0.9$     & Spectrum shift: $+10$\% TeV tail   \\
Merger rate        & $24^{+14}_{-9}\;\mbox{Gpc}^{-3}\ \mbox{yr}^{-1}$ & Detection rate: $\sim$linear \\
\hline
\end{tabular}
\caption{Estimated sensitivity of predicted photon flux to key model parameters.}
\label{tab:uncertainty}
\end{table}

 \subsection{Supermassive BH mergers}

Several SMBH binary systems are known in galaxies, probably following the merger of two progenitor galaxies which hosted a SMBH at their centre. It is believed that the galaxy merger generates a binary system at distances down to 1~pc, while further decay leading to a merger is due to the interactions with the star population \cite{Milosavljevic:2002ht}. Such merger would produce gravitational waves in the sensitivity ranges of Pulsar timing array telescopes (PTA) \cite{Hobbs_2010} or LISA.  Intriguingly, it has recently been reported that the activity of the galaxy SDSS J1430+2303, at a distance of $378$~Mpc from Earth hints at a possible SMBH merger within three years \cite{Jiang:2022aek,Dou:2022atu}. The system is currently being monitored by HESS \cite{HESS:2023otq}, while a gravitational wave signal will be in the observation window of PTAs.

\begin{figure}[htb]
\centering
\includegraphics[width=8cm]{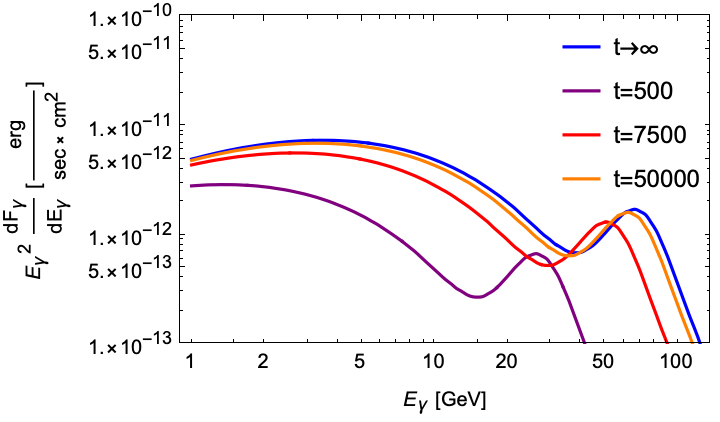} 
\caption{Differential flux on Earth from one solar mass of BH morsels of individual masses $10^9$kg at a distance of $300$ Mpc. We assume that the morsels were produced at a distance $0.5$ Schwarzschild radius from the horizon of the merged SMBH, and with energy $\frac{E}{M_\text{Bm} c^2}=1+10^{-7}$. The spectrum is heavily distorted at initial times by the gravitational field of the merged SMBH of mass $10^8\ M_\odot$.}
\label{Fig:SMBH}
\end{figure}

Because of the much larger masses, a SMBH merger can potentially emit much more mass in BH morsels, hence providing more intense GRBs than a stellar mass merger. Furthermore, the initial photons emitted by heavier morsels could be visible and provide steady signals that would outlive the photons emitted by other astrophysical sources. Another interesting feature is the fact that the BH morsels are produced and evaporate within the strong gravitational field of the merged SMBH, hence their spectrum may record a signature of where the BH morsels have been produced and the mass of the merged SMBH. For illustration purposes, and for simplicity, we computed the gravitational effects considering a Schwarzschild SMBH of mass $10^8\; M_\odot$ and a population of morsels of mass $10^9$~kg produced at a given distance and energy. We include the main effects due to the time dilation and energy redshift induced by the SMBH  gravity. In Fig.~\ref{Fig:SMBH}, we show the expected differential flux at different times for BH morsels produced at a distance of $0.5$ Schwarzschild radius from the SMBH horizon. At early times, the flux is reduced and redshifted, while it approaches the asymptotic spectrum (blue curve) after about $50.000$~sec. After that, the emission remains constant until the evaporation time is reached. The timing of the effect depends crucially on the initial energy of the BH morsels at production.

\subsection{Outlook}

So far, we have presented simple templates based on BH morsels of the same mass emitted radially as a spherically symmetric shell. An extended mass distribution can modulate and prolong the duration of the TeV photon emission. The geometry of the emission, instead, has a smaller impact on the signal, as BHs emit particles in a spherically symmetric manner. Hence, as long as the morsels are not screened by the merged BH, their Hawking radiation will reach our detectors. Some distortions of the spectrum can be expected if the BH morsels travel with relativistic velocity, in which case their direction of motion with respect to the line of sight becomes important. When BH morsels are emitted in a strong gravitational field, such as for SMBH mergers, their Hawking emission may be affected and a dedicated study is required to quantify such effects, which we leave for future investigations. As a final remark, we recall that BH morsels will most intensely emit at times close to their evaporation and with primary energies exceeding the multi-TeV scale. Besides a correct characterisation of the emission of secondary spectra \cite{Bauer:2020jay}, the Hawking radiation at such energies can be impacted by the limited validity of the standard model of particle physics, and some new physics effects may induce measurable distortions \cite{Baker:2022rkn,Federico:2024fyt}. The observation of Hawking radiation from BH morsels, therefore, could enlighten us not only about the production of such morsels, but also about particle physics at energies beyond the reach of current and future collider experiments, carrying imprints from new physics based on supersymmetry, composite dynamics, or extra dimensions, to name a few.

BH morsels are not the only source of particles being emitted during a merger of compact objects. 
Several astrophysical mechanisms can produce high-energy transients in the vicinity of BH mergers, including magnetospheric reconnection~\cite{Lyutikov:2011vca, Kelly:2017xck}, accretion disk disruption~\cite{Rosswog:2005su, Metzger:2011bv}, and jet activity~\cite{LIGOScientific:2017ync, Ajello:2019avs}.  Morsel evaporation events are characterized by a delayed, isotropic, multi-TeV burst coincident with a GW event, lacking persistent afterglow emission or strong beaming. In contrast, disk- or jet-driven GRBs often show beamed emission, multiwavelength afterglows, and may not correlate tightly in time with GW triggers.

\section{Conclusions}

In this work, we  have shown that it is possible to observe the Hawking radiation emitted by BH morsels assumed to form in catastrophic astrophysical events such as astrophysical BH mergers. The resulting GRBs offer unique footprints of these asteroid-mass BH morsels emitted after the merger, with photon energies exceeding the TeV scale. The time delay between the gravitational wave event and the GRB is correlated to the mass distribution of the morsels. We showed that the total morsel mass allowed by the unaccounted merger mass leads to a Hawking induced radiation in photons that is above the sensitivity of atmospheric Cherenkov telescopes such as HESS, LHAASO and HAWC, and that current data already constraints the total mass emitted in morsels to the level of a thousandth solar mass.

\vspace{0.2cm}

\section*{Acknowledgements}
We thank Alexandre Arbey for help and assistance with BlackHawk.
We also thank Zhi-Wei Wang and Juri Smirnov for inspiring discussion at the beginning of this project. The work of F.S. is
partially supported by the Carlsberg Foundation, semper ardens grant CF22-0922.

\appendix
\section{Morsel Creation in Theories beyond GR}\label{App:Motivation}

Colliding BHs have been identified since long time ago as a source of gravitational waves \cite{Hawking:1971tu}, with the expected outcome of a single merged BH. This expectation follows from a simple application of the area law \cite{Bekenstein:1972tm,Bekenstein:1973ur} in GR: while multiple BH final states are thermodynamically allowed, the solution with a single merged one is entropically favoured. However, theories beyond GR have been developed,  where fragmentation of black holes occurs. Even within GR, considerations regarding the topology of the horizon typically refer to static solutions of Einstein's equations \cite{Israel:1967za,Hawking:1971vc}: BH mergers are definitely more complicated solutions. Indeed, numerical simulations within GR have already shown the emergence of strong non-linearities in the gravitational field both during the merger \cite{Okounkova:2020vwu}  and during the successive ringdown phase \cite{Giesler:2019uxc,Khera:2023oyf}. Henceforth, one cannot exclude that the extreme environment occurring during the merger could give rise to the production of small BHs, that are successively expelled by the system. Note that non-linearities may also emerge in highly rotating BHs \cite{Yang:2014tla,Gralla:2016sxp,Westernacher-Schneider:2017xie,Bini:2022ewn}.

\subsection{Fragmentation of Black Holes}
The horizon topology of BHs in 4-dimensional GR is constrained to be spherical \cite{Israel:1967za,Hawking:1971vc} and the horizon is resilient to perturbations, rendering the fragmentation of a fully formed horizon unlikely. The situation changes in some extension of general relativity, and we present a non-exhaustive list of possibilities below:

\begin{itemize}

    \item In extra dimensional space-time, BH solutions are less restricted. This leads to geometries that exhibit instabilities, even classically, and thus may decay and fragment. For example, it was argued \cite{Gregory:1993vy,Gregory:1994bj} that black strings, \emph{i.e.} metrics with a horizon geometry of the form $S^2\times S^1$, and more generally black branes, are unstable under small perturbations. Furthermore, higher dimensional BH solutions are known (\emph{e.g. }\cite{MYERS1986304}) that allow for arbitrarily high angular momentum per unit mass (thus avoiding the Kerr bound in 4-dimensions \cite{Kerr:1963ud}). It was argued in \cite{Emparan:2003sy} that such ultra-spinning BHs exhibit instabilities beyond a certain threshold and thus may also fragment. Such characteristics may survive in models with compact extra dimensions where only gravity propagates \cite{Arkani-Hamed:1998jmv}, hence the products of the horizon fragmentation would resemble BH morsels from a 4-dimensional perspective. In such a scenario, the size of the extra dimensions, which is bound to be below a few microns, would provide a characteristic scale for the morsels.

    \item More general higher dimensional configurations have been studied in the context of superstring theory. The spectrum of these BPS solitons (\emph{e.g.} \cite{Denef:2000nb}) generically depends on the position in the moduli space of the theory: across certain co-dimension one surfaces (called walls of marginal stability), single-particle BPS states, resembling (charged) BHs from a 4-dimensional perspective, may decay into more elementary multi-centre configurations. For more information on this wall-crossing phenomenon, we refer the reader to the review article \cite{Pioline:2011gf} and references therein. We simply remark that string theory generally allows the fragmentation of solitonic solutions, which may be interpreted as the formation of BH morsels.

    \item  BH horizon instabilities and fragmentation have also been investigated in 4-dimensional modifications of general relativity. For example, in \cite{Chen:2017kpf} it was shown via   entropy arguments that combining gravity extensions of the form f(R) \cite{Capozziello:2011et} (including monopole dynamics) with the modified uncertainty principle stemming from the existence of a minimum length \cite{Amati:1988tn,Kempf:1994su,Kempf:1996nk} leads to BH instabilities and fragmentation.

\end{itemize}
Furthermore, we remark that morsels may also be produced thanks to the presence of matter or fields around the BH horizon. For instance, in the presence of ultra-light bosonic degrees of freedom, such as axion-like dark matter candidates, a Bose condensate forms near the BH horizon via superradiance \cite{Brito:2015oca}. It has been shown that, around rotating Kerr BHs, the value of the field can reach values close to the Planck scale \cite{Chen:2022kzv}. This scenario, therefore, predicts the formation of a shell enveloping the BH, which has been recently considered as a source of accelerated fermions \cite{Chen:2023vkq}.   During a BH merger, the condensate shell may be trapped in between the colliding horizons, be compressed and locally collapse into BH morsels. Note that other effects of superradiance on BH mergers have been studied in \cite{Payne:2021ahy,Aurrekoetxea:2023jwk,Tomaselli:2024dbw}.
A scenario involving the Higgs field was considered in \cite{Chitishvili:2021jrx}, under the hypothesis that the strong gravitational field near the horizon modifies the Higgs potential and generates a new vacuum configuration. Hence, bubbles of the new vacuum form near the horizon. During a merger, for a short time, the bubbles are trapped in between the merging horizons and collide, hence producing small BHs. 
    
Finally, we would like to mention the effect of a very light scalar field, with Compton wavelength of the same order as the astrophysical BH radius. Such light states may emerge as axion-like particles in many extensions of the standard model of particle physics. It has been shown that a high energy density can accrete in such scalar field around BHs with \cite{Chen:2022kzv} or without spin \cite{Barranco:2012qs,Barranco:2013rua,Aguilar-Nieto:2022jio}. Effectively, this effect would create an envelope around the horizons of the merging BHs at a distance of a few radii and with densities of the order of the BH density \cite{Aguilar-Nieto:2022jio}. During the merger, the collision of such envelopes could produce morsels, as long as a strong compression is generated during the merger by shock waves of the strong non-linearity in the gravitational field. A detailed study of such mechanism, however, is beyond the scope of this work, and also out of reach from numerical simulation due to the small scales involved. Note on the side that the presence of scalar envelopes could also affect the spectrum of gravitational waves, hence providing a second characteristic signature \cite{Payne:2021ahy,Aurrekoetxea:2023jwk,Tomaselli:2024dbw}. 

We leave a detailed study of the production mechanisms for the morsels to a future
publication, as the main focus of the present work is to establish the potential observability
of the morsels via their Hawking radiation.


\subsection{Back-of-the-Envelop Estimation of Morsel Mass}\label{Sect:Estimation}

A different point of view that naively allows for the fragmentation of BHs (particularly in a quantum theoretical setting) stems from the central dogma \cite{Strominger:1996sh}, which states that a BH can be described in terms of an ensemble of quantum states, to construct a toy model leading to the production of morsels. Hence, we model the BH as a bound state of internal components, which follow an appropriate equation of state that balances the gravitational force and prevents further collapse. Qualitatively,  this description is fairly similar to that of neutron stars (NSs): it is well known that during NS mergers, a large number of nucleons (the inner components of the NS) are emitted. Numerical simulations estimate that, depending on the conditions of the merger, the total mass ejected in debris can amount up to $10^{-3}$ or $10^{-1}$ solar masses \cite{Shibata:2019wef}. Stiffer equations of state lead to lesser ejected mass, hence for BH mergers we take conservatively $10^{-3}$ solar masses as a reasonable reference value. To estimate the mass of the ejected BH debris, which we identify with morsels, we need to model the inside of the astrophysical BH. Assuming for simplicity that the mass is uniformly distributed, with density $\rho_{\rm BH} = \frac{3 M}{4 \pi R^3}$, where $R = 2 G M$ is the BH radius, applying the inner Schwarzschild solution we find that the inner gravitational pressure is constant and equal to
\begin{equation}
    p_{\rm BH} = - \rho_{\rm BH} = - \frac{3}{8 \pi G R}\,.
\end{equation}
Interestingly, this follows the equation of state of vacuum energy. Hence, the interaction among the inner components, which stabilises the BH, would have a positive vacuum energy $V_{\rm int} = - p_{\rm BH}$. From this, we can estimate the typical interaction length:
\begin{equation}
    l_{\rm int} \sim \frac{1}{V_{\rm int}^{1/4}} = \left(\frac{8\pi}{3}\right)^{1/4} \sqrt{l_{\rm Pl}\ R}  = 6.8 \times 10^{-16}~\mbox{m} \times \sqrt{\frac{R}{10~\mbox{km}}}\,,
\end{equation}
where $l_{\rm Pl}$ is the Planck length and we used a typical radius of a stellar mass BH. Assuming that the emitted morsels have such a radius, their typical mass would be
\begin{equation}
    M_{\rm Bm} \sim  4.7 \times 10^{11}~\mbox{kg} \times \sqrt{\frac{R}{10~\mbox{km}}}\,.
\end{equation}
We can consider this as an order of magnitude estimate of the emitted morsels. Clearly, this is a simple minded model, which would need great refinement to properly describe a BH, however it allows us to estimate the key quantities for the study of the morsel phenomenology and observability. A more quantitative study would require numerical simulations of the merger with an incredible precision to resolve the morsels, and with the unknown of the BH equation of state.

\bibliographystyle{utphys}
\bibliography{sample}

\end{document}